\documentclass[12pt,notitlepage,a4paper]{article}
\pdfoutput=1
\usepackage[a4paper,margin=1in]{geometry}
\usepackage{amsmath,amssymb,amscd,amsfonts}
\usepackage{mathtools}
\numberwithin{equation}{section}
\usepackage{url}
\usepackage[shortlabels]{enumitem}
\usepackage{tikz} 
\usetikzlibrary{decorations}
\usetikzlibrary{arrows,decorations.markings}         
\usepackage[hidelinks]{hyperref}
\usepackage[open,
  openlevel=2,
  atend,
  numbered]{bookmark}
\usepackage{graphicx}
\usepackage[final]{pdfpages}
\usepackage{threeparttable}
\usepackage[small,sc]{caption}
\usepackage{float}
\usepackage{stackengine}
\usepackage{scalerel}
\usepackage[bf,small,center]{titlesec}
\usepackage{tocloft}

\newlength{\mylen}	
\urldef\footurlb\url{cocalc.com/dfriedan/DM/SM}
\urldef\footurla\url{physics.rutgers.edu/~friedan}
\usepackage{sidecap}
\usepackage{cite}
\def\eq{\begin{equation}}
\def\en{\end{equation}}
\def\eqg{\eq\begin{gathered}}
\def\eng{\end{gathered}\en}
\def\eqa{\eq\begin{aligned}}
\def\ena{\end{aligned}\en}
\def\Reals{\mathbb{R}}

\DeclareMathOperator{\cn}{cn}

\def\expval#1{\langle \, #1 \,\rangle}

\def\gauge{\mathrm{gauge}}

\def\tr{\mathrm{tr}}
\def\Complexes{\mathbb{C}}

\def\vH{v}

\def\Higgs{\scriptscriptstyle \mathrm{Higgs}}

\def\CGF{\scriptscriptstyle \mathrm{CGF}}

\def\CDM{\scriptscriptstyle \mathrm{CDM}}

\def\GeV{\text{\footnotesize GeV}}
\def\sunit{\text{s}}
\def\ntimes{\,{\times}\,}

\def\gtwo{g}
\def\lambdaH{\lambda}

\stackMath
\def\dyhat{-0.2ex}
\newcommand\myhat[1]{\ThisStyle{%
              \stackon[\dyhat]{\SavedStyle#1}
                              {\SavedStyle\hat{\phantom{#1}}}}}
\def\that{\kern0.1em\myhat{\kern-0.1em t}}

\def\munit{\mathrm{m}}
\def\cmunit{\text{cm}}

\def\Msun{\textup{M}_{\odot}}
\def\texttilde{\raise-0.7ex\hbox{\!\texttt{\char`\~}}}
\def\scalar{\mathrm{scalar}}

\def\Planck{\scriptscriptstyle \mathrm{Planck}}

\def\aM{\ell}
\def\az{\ell_{0}}
\def\kgunit{\text{kg}}

\def\omegaz{\omega_{0}}

\def\now{\text{now}}
\def\central{\text{central}}
\def\rhoz{\rho_{0}}
\def\aone{a_{\ell}}
\titlespacing{\section}{3pc}{1.75pc}{0.8pc}
\titleformat{\section}
  {\centering\normalsize\bfseries} 
  {\large\thesection}
  {0.5em} 
  {\large}
\titleformat{\subsection}
  {\normalsize\bfseries} 
  {\thesubsection}
  {0.5em} 
  {}
\begin{document}
\def\title{The CGF dark matter fluid}
%

\begin{center}
{\Large \title}
\vskip5ex
{\large Daniel Friedan}
\vskip2ex
{\it
New High Energy Theory Center
and Department of Physics and Astronomy,\\
Rutgers, The State University of New Jersey,\\
Piscataway, New Jersey 08854-8019 U.S.A.
\vskip0.5ex
\href{mailto:dfriedan@gmail.com}{dfriedan@gmail.com}
\qquad
\href{https://physics.rutgers.edu/\textasciitilde friedan/}
{physics.rutgers.edu/\texttilde friedan}
}
\vskip2ex
\today
\end{center}
%
%
\begin{center}
\vskip3ex
{\sc Abstract}
\vskip3ex
\parbox{0.96\linewidth}{
\hspace*{1.5em}
The cosmological gauge field (CGF) is a classical solution of
SU(2)-weak gauge theory oscillating rapidly in time.  It is the dark
matter driving the CGF cosmology.  A general, local, mathematically
natural construction of the CGF is given here.  The macroscopic
properties are derived.  The CGF is an irrotational perfect fluid.  It
provides a synchronized global time coordinate and a global rest
frame.  There is a conserved number density.  The energy density and
pressure are related by the same equation of state as derived in the
CGF cosmology and used in the TOV stellar structure equations for
stars made of CGF dark matter.  The present construction justifies the
TOV solution.  Some possible routes towards testing the theory are
suggested at the end.
}
\end{center}


%
%
%
\begin{center}
\tableofcontents
\end{center}
%
%

\section{Introduction}

\subsection{The CGF cosmology and dark matter stars}

If dark matter can be explained within the Standard Model then,
as far as we know,
the Standard Model (including some neutrino couplings) and classical
General Relativity (including the cosmological constant) are
the complete fundamental laws of physics which have governed the universe over all of its
history since a time before the electroweak transition.  A fundamental or first principles
cosmology is an initial state of the SM and classical GR at the
beginning of this Standard Model epoch.  Such an initial state completely determines
all following cosmology.

The CGF cosmology 
\cite{Friedan:2020poe,Friedan2022:Atheory,Friedan2022:Stability,Friedan2022:DMStars,Friedan2022:FirstPrinciples}
is a fundamental cosmology of the Standard  Model epoch.
It starts from a specific initial state
prior to the electroweak transition.
The initial state is semi-classical,
a classical solution of the SM and GR
corrected by small fluctuations.
The initial state is completely determined by a certain
Spin(4) symmetry group
and a large initial energy.
The Spin(4) symmetry acts as SO(4) on space, which is a 3-sphere,
imposing homogeneity and isotropy.
The Spin(4) acts nontrivially on the SU(2)-weak sector of the SM.
The Spin(4)-symmetric SU(2)-weak gauge field has a single 
degree of freedom $b(t)$.
This is the cosmological gauge field (the CGF).
$b(t)$ is an anharmonic oscillator
because the Yang-Mills hamiltonian is quartic.
$b(t)$ oscillates rapidly in time because of its high initial energy.

The macroscopic energy-momentum tensor of the CGF is that of perfect fluid which is
non-relativistic at low density ($w\approx 0$).
Its non-gravitational interactions are very small.
So the CGF is dark matter.  
At leading order, in the classical approximation, the universe
contains only the CGF dark matter. 
Ordinary matter is a sub-leading correction arising
from fluctuations around the classical solution.
There is a systematic expansion around the classical dark matter 
universe.

This simple initial condition 
leads to an electro-weak transition
followed by an expanding universe that contains primarily dark matter.
The dark matter is a coherent effect within the Standard Model.
No physics beyond the Standard Model is needed.

The initial energy is the only free parameter of the classical
solution, but the specific value of the initial energy
does not affect
the local cosmology at all, so there are effectively no free
parameters.  The initial energy determines the radius of curvature, so
the  large lower bound on the radius of curvature from observation
puts a large lower bound on the initial
energy.

\begin{SCfigure}[][h]
\includegraphics[scale=0.6]{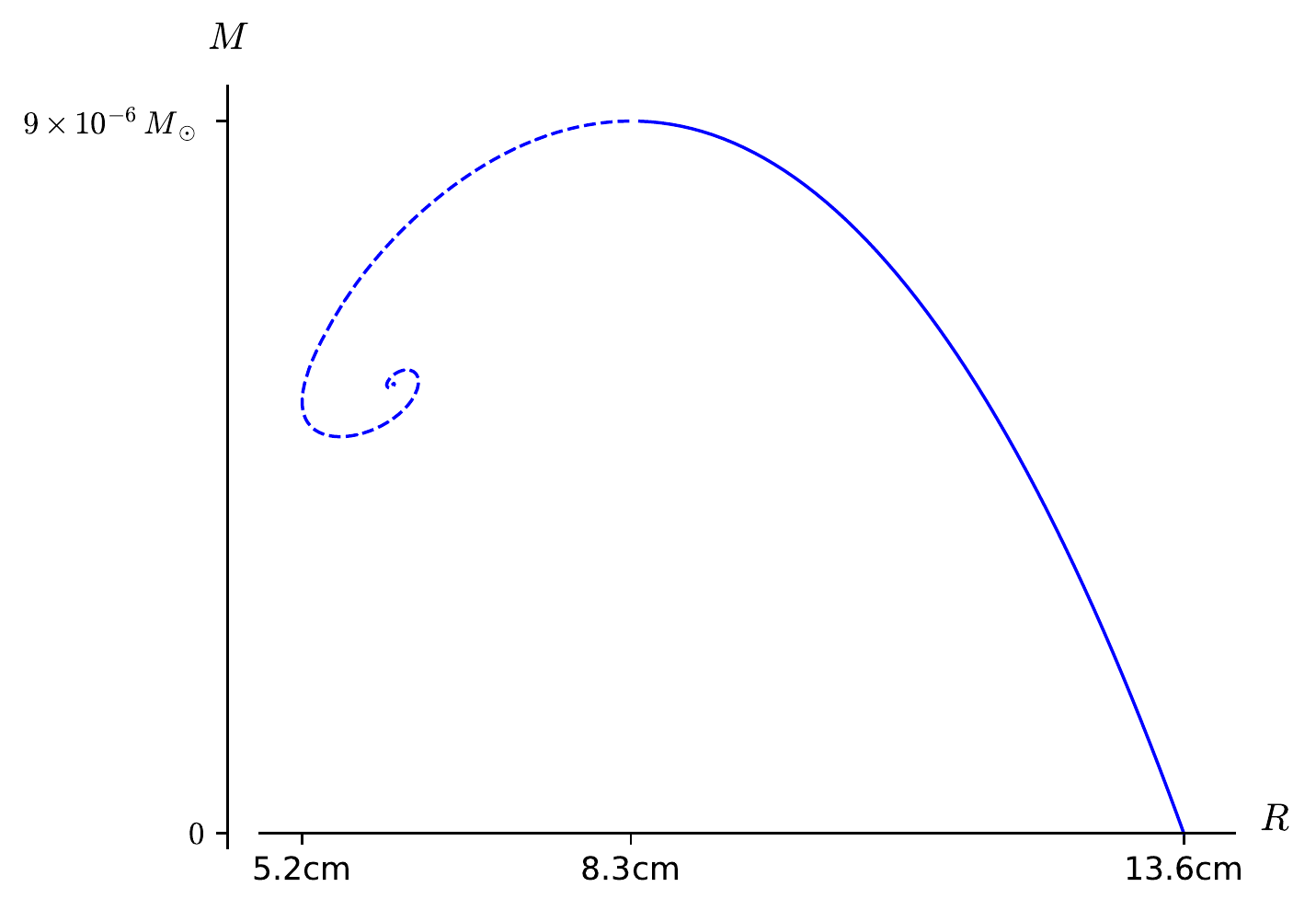}
\caption{
The curve of solutions of the TOV equations.
The dashed curve are the solutions unstable against spherically 
symmetric pulsations.
The solid curve are the (presumably) stable solutions.
If the dark matter
 takes the form of such compact objects,
microlensing observations imply that most of the dark matter
must lie in objects of low mass $M<10^{-11}\,\Msun$.
Such low mass CGF stars all have radius $13.6$\,cm
independent of the mass.
\vspace*{1ex}
\label{fig:MR}
}
\end{SCfigure}

The
classical solution $b(t)$ is given by an elliptic function periodic
in imaginary time as well as real time, thus defining a natural temperature.
The initial fluctuations of the SM fields around the classical CGF are in
the thermal state determined by the Spin(4) symmetry.
To check the feasibility of
the CGF cosmology, it is crucial to calculate the time evolution of
the initial fluctuations around the classical CGF.
Overdensities in the CGF have presumably collapsed to 
self-gravitating bodies.
The TOV stellar structure equations
for stars made of the CGF dark matter fluid
were solved in \cite{Friedan2022:DMStars}.
The solutions are shown in Figure~\ref{fig:MR}.
If the CGF cosmology is correct
then presumably
the dark matter is now in the form of such dark matter stars.

\subsection{Summary of the construction}

The solution of the TOV equations in \cite{Friedan2022:DMStars} assumed that the
non-homogeneous CGF behaves as a perfect fluid
obeying the same equation of state
as derived in the homogeneous, isotropic cosmology.
The assumption is justified here
by 
a general, local microscopic construction of the classical CGF
in terms of the SM fields.
This general construction
can be used to calculate
the time evolution of the density fluctuations
in order to determine if and when the fluctuations end up as
halos consisting of dark matter stars.
The subleading corrections to the classical CGF can be calculated,
in particular to find whether the interactions  with the SM fields 
are strong enough
to be used to detect the CGF.

The general CGF is
a solution of the SU(2)-weak gauge field equation of motion
which oscillates
in time on the microscopic scale
\eq
\az = \frac{\hbar}{m_{\Higgs}} = 5\ntimes 10^{-27}\sunit
\qquad
\text{(in $c=1$ units)}
\en
while varying in space at a much larger macroscopic scale $\aM$.
The space-time metric is
\eqg
ds^{2} =\aM^{2} \,g_{\mu\nu}(x)\,dx^{\mu}dx^{\nu}
\qquad
\text{signature } (-,+,+,+)
\eng
where $g_{\mu\nu}(x)$ and $x^{\mu}$ are dimensionless quantities.
Indices are raised and lowered with $g_{\mu\nu}$.
Powers of the macroscopic length $\aM$ are written explicitly.
$\aM$ is determined by the 
Einstein equation:
\eqg
G_{\mu\nu} = 8\pi G  \,{\aM^{2}}g_{\mu\sigma} T^{\sigma}_{\nu}
\qquad
G_{\mu\nu} = O(1)
\qquad
T^{\sigma}_{\nu}
= O\left(\frac{m_{\Higgs}}{\az^{3}}\right)
\\[1ex]
1 =  G  \aM^{2}\frac{m_{\Higgs}}{\az^{3}}
= \frac{G  \aM^{2}m_{\Higgs}^{4}}{\hbar^{3}}
\qquad
\aM = \frac{\hbar }{m_{\Higgs}^{2}}\sqrt{\frac{\hbar}{G}}
= \frac{\hbar m_{\Planck}}{m_{\Higgs}^{2}}
 = 15.4\,\cmunit
\eng
The large ratio
\eq
\omegaz = \frac{\aM}{\az} 
=  \frac{m_{\Planck}}{m_{\Higgs}}
=  1\ntimes 10^{17}
\en
suppresses spatial derivatives in the energy-momentum tensor
relative to time derivatives,
which is to say that
the hamiltonian is ultralocal
up to $O(10^{-17})$  corrections.

The logic of the construction is:
\begin{enumerate}
\item
The construction should
be local in space-time,
work in an arbitrary space-time metric,
and be mathematically natural (generally 
covariant).

\item
The CGF cosmology
should be the special case
with Spin(4) symmetry.

\item
The rank-2 SU(2)-weak vector bundle
can be identified with the rank-2 vector bundle of chiral spinors on 
space-time
up to an arbitrary SU(2) gauge transformation.
The gauge fields
in the spinor bundle become the SU(2)-weak gauge 
fields.
The general such identification is parametrized by
a field $u^{\mu}(x)$ of time-like unit 4-vectors,
$u^{\mu}u_{\mu} = -1$, $u^{0}>0$.

\item
There is a mathematically natural family of SU(2) gauge fields in the spinor 
bundle parametrized by a single scalar function $b(x)$ on space-time.

\item
The microscopic CGF solves the Yang-Mills equation in this 
natural family of gauge fields
on the condition of rapid oscillation in time.
It is a mathematically natural classical solution of the SU(2)-weak gauge theory,
well-defined up to gauge equivalence.

\item
The 3d wavefronts provide a global synchronized time coordinate $T(x)$
and a global rest frame.

\end{enumerate}

\subsection{Summary of the macroscopic properties}

The macroscopic energy-momentum tensor is that of a perfect fluid 
of density $\rho(x)$, pressure $p(x)$,
and four-velocity $u_{\alpha}(x)$,
\eq
T^{\alpha}_{\mu} =  -  u^{\alpha}u_{\mu} \rho + 
(\delta^{\alpha}_{\mu} + u^{\alpha}u_{\mu} )p
\en
There is a number density $n(x)$
that obeys a continuity equation,
\eq
\nabla_{\mu} J^{\mu} = 0
\qquad
J^{\mu}(x) = n(x) u^{\mu}(x)
\en
The four-velocity is irrotational,
\eq
u_{[\mu}\partial_{\nu} u_{\rho]} = 0
\en
Irrotationality is equivalent to existence of the global rest frame
in which $u^{i}(x)=0$ and $g_{0i}(x) =0$.

The CGF has two phases.  At high density the SU(2) 
gauge symmetry is unbroken.
At low density it is broken.
In the broken phase
the CGF is parametrized by a numerical function $k(x)$.
Density, pressure, and number density 
are given by parametric equations
\eq
\label{eq:parametricequationsintro}
k \mapsto \rho(k)  \qquad
k \mapsto p(k) 
\qquad
k \mapsto n(k) 
\en
which are algebraic expressions in
the complete elliptical integrals $K(k)$ and $E(k)$ of the first and second kinds.
$k(x)$ is the elliptic modulus that parametrizes the anharmonic 
oscillation of $b(x)$.
The equation of state relating density and pressure is
given implicitly by the parametric equations 
(\ref{eq:parametricequationsintro}).
It is the same equation of state as derived in the cosmological construction \cite{Friedan2022:Atheory}
and used in the TOV equations for CGF dark matter stars \cite{Friedan2022:DMStars}.

Details of calculations are shown in a separate note \cite{Friedan2022:CGFFluidSuppMat}.

\section{Standard Model action}

The SM fields that participate in the CGF are 
the SU(2)-weak gauge field $B_{\mu}(x)$
and the SU(2) doublet Higgs field $\phi(x)$.
The SU(2) covariant derivative is 
\eq
D_{\mu} = \partial_{\mu}+B_{\mu}
\qquad
D_{\mu}\phi = \partial_{\mu}\phi+B_{\mu}\phi
\en
The curvature form is
\eq
F_{\mu\nu} = [D_{\mu},\,D_{\nu}]
= \partial_{\mu}B_{\nu}  - \partial_{\nu}B_{\mu} +[B_{\mu},\,B_{\nu}]
\en
The classical action as parametrized in \cite{PDG} is
\begin{align}
\label{eq:gaugeaction}
\frac1\hbar S_{\gauge}
&=\int  \frac1{2 g ^{2}}
\tr\left(-\aM^{-4} F_{\mu\nu} F^{\mu\nu}\right)  \,\aM^{4} \sqrt{-g} \,d^{4} x\
\\[1ex]
\label{eq:scalaraction}
\frac1\hbar S_{\scalar}  
&=\int \left[\aM^{-2}D_{\mu}\phi^{\dagger} D^{\mu}\phi
+ \frac{1}{2} \lambdaH^{2}\left(\phi^{\dagger}\phi 
-   \frac12 \vH^{2}\right)^{2}\right]\, \aM^{4}\sqrt{-g} \,d^{4} x
\end{align}

\eqg
\label{eq:parametermassrelations}
g = \frac{e}{\sin \theta_{W}}
\qquad
\cos \theta_{W}  = \frac{m_{W}}{m_{Z}}
\qquad
\frac{g v}{2} = \frac{m_{W}}{\hbar}
\qquad
\lambda v = \frac{m_{\Higgs}}{\hbar}
\eng

\eqg
\alpha = \frac{e^{2}}{4\pi} = 1/137.035999139(31)
\qquad
\frac{m_{W}}{m_{Z}} = 0.88147(13)
\\[1ex]
m_{W} = 80.379(12)\,\GeV
\qquad
m_{Z} = 91.1876(21) \,\GeV
\qquad
m_{\Higgs} = 125.10(14)\,\GeV
\eng
giving tree-level parameters
\eq
\label{eq:treelevelparameters}
g^{2} =0.411
\qquad
\lambda^{2} = 0.249
\en
The CGF probes energies on the order of $m_{\Higgs}$
but 3-momenta on the macroscopic scale $\hbar/\aM$
so there should be no appreciable renormalization of the
couplings.

\section{SU(2) structure on space-time spinors}

\subsection{Equivalence to a four-velocity field $u^{\alpha}(x)$}

Let $S$ be the vector bundle of
spinors over space-time.
The fiber $S_{x}$ at $x$ is a four-dimensional complex vector space.
Let $S^{+}$ be the rank-two sub-bundle of chiral 
spinors (the eigenspaces $\gamma_{5}=1$).
The fiber $S^{+}_{x}$
at each point $x$ in space-time is the 
two-dimensional 
defining representation of the group Spin(1,3) = SL(2,$\Complexes$).
An SU(2) structure on the chiral 
spinors is an SU(2) subgroup of SL(2,$\Complexes$)
determined by a positive hermitian form on $S^{+}_{x}$.
That is,
$
\text{SU}(2) =  \text{SL}(2,\Complexes) \cap \text{U}(2)
$.
The space of SU(2) structures on $S^{+}_{x}$ is
\eq
\text{SL(2,$\Complexes$)}/\text{SU(2)} = \text{SO(1,3)}/\text{SO(3)} = \{ u^{\alpha}\colon
u^{\alpha}u_{\alpha} = -1, \; u^{0} > 0\}
\en
i.e., the space of four-velocities $u^{\alpha}(x)$ at $x$.
The SU(2) structures on the vector bundle $S^{+}$
are in 1-to-1 correspondence with the
four-velocity fields $u^{\alpha}(x)$.

Given an SU(2) structure on the chiral spinors,
i.e., given a four-velocity field $u^{\alpha}(x)$,
the SU(2)-weak vector bundle
can be identified with
$S^{+}$
up to gauge equivalence.
Conversely, any such identification determines an SU(2) structure on 
the chiral spinors.

The CGF is
a natural solution of the Yang-Mills equation 
in the SU(2) spinor bundle
arising from such an identification.
The four-velocity field $u^{\alpha}(x)$ characterizing the SU(2) structure
turns out to be the physical four-velocity field of the macroscopic CGF fluid.

The construction is local in space-time.  It extends to a global 
construction if no topological obstructions
prevent
identifying the SU(2)-weak vector bundle with the chiral spinor 
bundle.
There are no such obstructions
in the space-time  $\Reals \times S^{3}$ of the CGF cosmology.

\subsection{A natural SU(2) spin connection}

The space-time metric is
\eq
ds^{2}  =\aM^{2} \, g_{\mu\nu}(x)\, dx^{\mu}dx^{\nu}
\qquad
\text{signature } (-,+,+,+)
\en
with $\aM = 15.4\,\cmunit$ the macroscopic length scale.
The space-time orientation is expressed by
the volume form $\aM^{4} \epsilon_{\mu\nu\rho\sigma}(x)$.
\eq
\frac1{4!} \epsilon_{\mu\nu\rho\sigma}\epsilon^{\mu\nu\rho\sigma} = -1
\en
The  four-velocity field $ u^{\alpha}(x)$ expresses the SU(2) 
structure.
\eq
u_{\alpha}u^{\alpha} = -1 
\qquad
u^{0} > 0
\en
The projection on the space-like tangent vectors orthogonal to $u$ is
\eq
P^{\alpha}_{\beta} = \delta^{\alpha}_{\beta}  + u^{\alpha}u_{\beta}
\en
The Dirac matrices $\gamma_{\mu}(x)$ act on the spinors $S_{x}$ at 
$x$.
\eqg
\gamma_{\mu} \gamma_{\nu} +\gamma_{\nu} 
\gamma_{\mu}  = 2 g_{\mu \nu}
\qquad
\gamma_{5} = \frac{i}{4! }\epsilon^{\mu\nu\rho\sigma}
\gamma_{\mu} \gamma_{\nu} \gamma_{\rho} 
\gamma_{\sigma} 
\qquad
\gamma_{5}^{2} = 1
\eng
The generators of Spin(1,3) on $S_{x}$ are
\eq
L_{\mu\nu} = \frac14 [\gamma_{\mu},\,\gamma_{\nu}]
\qquad
L_{\mu\nu} = \frac1{2i} 
\epsilon_{\mu\nu}{}^{\rho\sigma}L_{\rho\sigma} \gamma_{5}
\qquad
[L_{\mu\nu} ,\,\gamma_{5}] =0
\en
The bundle of chiral spinors $S^{+}$  is the rank 2 subbundle of eigenspaces 
$\gamma_{5}(x)=1$.  The subbundle $S^{-}$ is the subbundle $\gamma_{5}(x)=-1$.
The positive hermitian form on $S^{+}_{x}$ associated to 
$u^{\alpha}(x)$ extends naturally to a 
positive hermitian form on $S^{-}_{x}$ and thus all of $S$ such that
\eq
u^{\alpha}\gamma_{\alpha}^{\dagger} = - u^{\alpha}\gamma_{\alpha}
\qquad
P^{\alpha}_{\beta}\gamma_{\alpha}^{\dagger} = P^{\alpha}_{\beta}\gamma_{\alpha}
\en
The boosts relative to $u^{\alpha}(x)$ are the matrices
\eq
L_{\mu} = u^{\alpha} L_{\alpha\mu}
\qquad
u^{\mu}L_{\mu} = 0
\qquad
L_{\mu}^{\dagger} = L_{\mu}
\en
The matrices $iL_{\mu}$ generate SU(2) on $S^{\pm}$.
\eq
[L_{\mu},\,L_{\nu}] 
= u^{\rho}\epsilon_{\rho\mu\nu}{}^{\sigma} i  L_{\sigma} \gamma_{5}
\qquad
L_{\mu}L_{\nu}+L_{\nu}L_{\mu} =\frac12 P_{\mu\nu}
\en
The metric covariant derivative 
$\nabla_{\mu}$ acts on spinors such that
\eq
\nabla_{\mu} \gamma_{\nu} = 0
\en
The curvature 2-form of $\nabla_{\mu}$ is 
\eq
F^{\text{metric}}_{\mu\nu} = [\nabla_{\mu},\, \nabla_{\nu}]
= \frac12 R^{\alpha\beta}{}_{\mu\nu}L_{\alpha\beta}
\en
$\nabla_{\mu} $ defines a Spin(1,3) connection in the spinor bundle.
It does not preserve the hermitian structure, i.e.~$\nabla_{\mu}\gamma_{\nu}^{\dagger} \ne 
\nabla_{\mu}\gamma_{\nu}$ or
$\nabla_{\mu}L_{\nu}^{\dagger} \ne 
\nabla_{\mu}L_{\nu}$.
The modified covariant derivative
\eq
D^{0}_{\mu} = \nabla_{\mu}  - \nabla_{\mu}u^{\sigma} L_{\sigma}
\en
defines a natural Spin(1,3) connection that does preserve the hermitian structure
\eq
D^{0}_{\mu} L_{\nu} = \nabla_{\mu}u^{\sigma} u_{\nu} L_{\sigma}
\qquad
D^{0}_{\mu} \left(L_{\nu}^{\dagger}\right) = (D^{0}_{\mu} L_{\nu})^{\dagger}
\en
so  $D^{0}_{\mu}$ defines a natural SU(2) connection in the spin bundle.  
Its curvature 2-form is
\eqg
F^{0}_{\mu\nu} =[D^{0}_{\mu},\,D^{0}_{\nu}]
= \left(\frac12 R^{\alpha\beta}{}_{\mu\nu} 
- \nabla_{\mu} u^{\alpha}\nabla_{\nu}u^{\beta} \right)
[L_{\alpha},\,L_{\beta}]
\eng

\section{Form of the CGF}

The general SU(2) covariant derivative is
\eq
D_{\mu}=D^{0}_{\mu}+B_{\mu}
\qquad
B_{\mu}= i L_{\alpha}B_{\mu}^{\alpha}(x) 
\en
The CGF has the naturally distinguished form
\eq
D_{\mu}=D^{0}_{\mu}+i  L_{\mu}b(x)
\en
parametrized by a scalar function $b(x)$ on space-time.
$b(x)$  oscillates in time on the microscopic scale
while varying smoothly in space on the macroscopic scale.
That is,
\eq
b(x) = \omegaz b_{0}(\zeta)
\qquad
\zeta = \omegaz T(x)
\en
where $b_{0}(\zeta)$ is a smooth function of $\zeta$
and also depends (implicitly for now)
smoothly on $x^{\mu}$
and where $T(x)$ is a smooth time coordinate,
\eq
\partial_{\mu}T\partial^{\mu}T < 0
\qquad
\partial_{0}T > 0
\en
We can take $b_{0}(\zeta)$ to be periodic in $\zeta$ with period 
$2\pi$ by reparametrizing $T(x)$.

Alternatively, the wave form of $b(x)$
determines $T(x)$.
Count wavefronts by consecutive integers $n$.
Then $T(x) = 2\pi n/\omegaz$ on the wavefronts
and interpolates smoothly between them.

\section{Equations of motion}

\paragraph{Scalar field equation of motion.}

Re-write the scalar action (\ref{eq:scalaraction})
in terms of the dimensionless scalar field $\hat \phi = 
{v\phi}/{\sqrt 2}$.
\eq
\frac1\hbar S_{\scalar}  
= \int \frac1{8\lambda^{2}} \left[\frac{4}{\omegaz^{2}} D_{\mu}\hat\phi^{\dagger} 
D^{\mu}\hat\phi
+ \left(\hat\phi^{\dagger}\hat\phi 
-   1\right)^{2}\right]\, \omegaz^{4}\sqrt{-g} \,d^{4} x
\en
with
\eqa
D_{\mu}\hat\phi^{\dagger} D^{\mu}\hat\phi &= 
(D^{0}_{\mu}+B_{\mu})\hat\phi^{\dagger} (D^{0\mu}+B^{\mu})\hat\phi 
\ena
$B_{\mu} B^{\mu}$ is a multiple of the identity, 
$B_{\mu} B^{\mu} = \frac12\tr (B_{\mu} B^{\mu})$,
so
\eqa
D_{\mu}\hat\phi^{\dagger} D^{\mu}\hat\phi &= 
D^{0}_{\mu}\hat\phi^{\dagger}D^{0\mu}\hat\phi 
- \hat\phi^{\dagger}B_{\mu}D^{0\mu}\hat\phi 
+D^{0}_{\mu}\hat\phi^{\dagger}B^{\mu}\hat\phi 
-
\hat\phi^{\dagger}\hat\phi \frac12\tr (B_{\mu} B^{\mu})
\ena
Assume $\hat\phi(x)$ is smooth and $B_{\mu} = O(\omegaz)$.
Then to leading order in $\omegaz$
\eq
\label{eq:scalaractionleading}
\frac1\hbar S_{\scalar}  
=
\int \frac1{8\lambda^{2}} \left[
2\hat\phi^{\dagger}\hat\phi
\, \tr \left(-
 \frac{B_{\mu} B^{\mu}}{\omegaz^{2}}\right)
+  \left(\hat\phi^{\dagger}\hat\phi 
-   1\right)^{2}\right]\, \omegaz^{4}\sqrt{-g} \,d^{4} x
\en
The CGF has  $B_{\mu} = \omegaz b_{0} i L_{\mu}$.
To leading order in $\omegaz$,
\eq
\frac1\hbar S_{\scalar}  
= \int \frac1{8\lambda^{2}} \left[
\,
(\hat\phi^{\dagger}\hat\phi)^{2}
+  3 \left (b_{0}^{2}-\frac23\right) \hat\phi^{\dagger}\hat\phi
+1
\right]\, \omegaz^{4}\sqrt{-g} \,d^{4} x
\en
Since $b_{0}$ is oscillating rapidly compared to the macroscopic scale,
$b_{0}^{2}$ can be replaced by its time average
$\expval{b_{0}^{2}}$.
The leading order scalar equations of motion are then

\eq
\begin{alignedat}{3}
\hat \phi &= 0 
\quad\; &
\expval{b_{0}^{2}} &\ge \frac23
&&\qquad\text{unbroken phase}
\\[1ex]
\hat\phi^{\dagger}\hat \phi &=    
1-\frac32 \expval{b_{0}^{2}}  
\quad\; &
\expval{b_{0}^{2}} &< \frac23 
&&\qquad\text{broken phase}
\end{alignedat}
\en
The solutions 
with $\expval{b_{0}^{2}} \ge \frac23$,
are in the {\it unbroken phase}.
The solutions 
with $\expval{b_{0}^{2}} > \frac23$,
are in the {\it broken phase}.
In the broken phase, when $\hat\phi^{\dagger}\hat \phi >0$,
the direction of $\hat \phi(x)$ is determined by the next-to-leading order 
equations of motion.

\paragraph{Gauge field equation of motion}

Vary the gauge action (\ref{eq:gaugeaction}) and the leading order scalar action
(\ref{eq:scalaractionleading}) wrt the gauge field
to obtain the gauge field equation of motion at leading order in 
$\omegaz$.
\eqg
\label{eq:gaugeeqnofmotion}
0 =  \frac{D^{\nu} F_{\mu\nu}}{\omegaz^{3}}
+\frac{g ^{2}\hat\phi^{\dagger}\hat\phi }{4\lambda^{2}}
\frac{B_{\mu}}{\omegaz}
+O(\omegaz^{-1})
\eng
where
\eqg
B_{\mu} = O(\omegaz)
\qquad
D^{0}_{\mu}B_{\nu} = O(\omegaz^{2})
\qquad
F_{\mu\nu} = D^{0}_{[\mu}B_{\nu]} + [B_{\mu},\,B_{\nu}] + O(\omegaz)
\eng
For the CGF,
\eqg
B_{\mu} = i b L_{\mu}
\qquad
b = \omegaz b_{0}(\zeta)
\qquad
\zeta = \omegaz T
\\[1ex]
\frac1{\omegaz^{2}} F_{\mu\nu}
= \frac{d b_{0}}{d\zeta} \partial_{[\mu}T  i L_{\nu]}
-b_{0}^{2} [L_{\mu},\,L_{\nu}]
+O(\omegaz^{-1})
\eng
The leading order equation of motion (\ref{eq:gaugeeqnofmotion}) is
\eqa
\label{eq:gaugeequationofmotionleading}
0 &= \left(-\partial^{\nu}T \partial_{\nu}T  \frac{d^{2} b_{0}}{d\zeta^{2}}
+\frac{g ^{2}\hat\phi^{\dagger}\hat\phi }{4\lambda^{2}}b_{0}
+2 b_{0}^{3}
\right) i L_{\mu}
+ \left(\frac{d^{2} b_{0}}{d\zeta^{2}} 
\partial_{\mu}T\partial^{\nu}T\right) i L_{\nu}
\\[1ex]
&\qquad \qquad{}
-\left(3 b_{0}\frac{d b_{0}}{d\zeta} \partial^{\nu}T\right) [L_{\mu},\, L_{\nu}]
\ena
Contracting with $u^{\mu}$ gives
\eq
0 = 
\left(\frac{d^{2} b_{0}}{d\zeta^{2}}  
u^{\mu}\partial_{\mu}T\partial^{\nu}T \right)  i L_{\nu}
\en
But $d^{2}b_{0}/d\zeta^{2}\ne 0$ and  $u^{\mu}\partial_{\mu}T\ne 0$ so $\partial^{\nu}T  L_{\nu}=0$ so 
\eq
\label{eq:kappa}
u_{\mu} = -\aone \partial_{\mu} T
\qquad
u^{\mu}\partial_{\mu}T = \frac1{\aone}
\qquad
\partial_{\nu}T \partial^{\nu}T = - \frac1{\aone^{2}}
\en
for $\aone(x)$ a dimensionless function on space-time.
Using (\ref{eq:kappa}) in (\ref{eq:gaugeequationofmotionleading}),
the leading order gauge field equation of motion 
becomes the anharmonic 
oscillator equation
\eq
0 = \frac1{\aone^{2}}   \frac{d^{2} b_{0}}{d\zeta^{2}}
+\left(\frac{g ^{2}\hat\phi^{\dagger}\hat\phi }{4\lambda^{2}}\right) b_{0}
+2 b_{0}^{3}
\en
which has conserved energy
\eq
\label{eq:conservedenergy}
H = \frac{1}2 \frac1{\aone^{2}}\left(\frac{d b_{0}}{d\zeta}\right)^{2}
+\frac12\left(\frac{g ^{2}\hat\phi^{\dagger}\hat\phi 
}{4\lambda^{2}}\right) b_{0}^{2}
+\frac12 b_{0}^{4}
\en
The oscillator is parametrized by $\aone^{2}$ and 
$\hat\phi^{\dagger}\hat \phi$ which vary slowly in space-time
compared to the period of oscillation.

\section{Solution of the equations of motion}
The Jacobi elliptic function $\cn(z,k)$  satisfies
\eq
1 
=\left(\frac{d \cn}{dz}\right)^{2}+(1-2k^{2}) \cn^{2} + k^{2}\cn^{4}
\en
and has period $4K$ in $z$,
where $K(k)$ is the complete elliptic integral of the first kind.
So
\eqg
F(\zeta,k) =  \frac{k\cn(z,k)}{\zeta'}
\qquad
\zeta = \frac{ 2\pi }{4K}z
\qquad
\zeta' = \frac{ 2\pi}{4K}
\eng
has period $2\pi$ in $\zeta$ and satisfies
\eq
\frac{k^{2}(1-k^{2}) }{2\zeta'{}^{4} }
 =\frac12\left(\frac{dF}{d\zeta}\right)^{2}+\frac12 
 \left(\frac{1-2k^{2}}{\zeta'{}^{2}} \right) F^{2}   + 
\frac12 F^{4} 
\en
The anharmonic energy equation (\ref{eq:conservedenergy}) is solved by
\eqg
b_{0}(\zeta) = \frac1{\aone} F(\zeta,k)
\eng
when $F$ satisfies
\eqg
\aone^{4}{H} =  \frac12 \left(\frac{dF}{d\zeta}\right)^{2}
+\frac12 \left( \frac{g ^{2}\hat\phi^{\dagger}\hat\phi\,\aone^{2}
}{4\lambda^{2}} \right) F^{2}
+\frac12 F^{4}
\eng
which is
\eqg
\label{eq:parametricequations}
H= \frac{ k^{2}(1-k^{2}) }{2\zeta'{}^{4}\aone^{4} }
\qquad
\frac{g ^{2} \hat\phi^{\dagger}\hat\phi}{4\lambda^{2}}  = 
\frac{1-2k^{2}}{\zeta'{}^{2}\aone^{2}}
\eng
The time average of $b_{0}^{2}$ is
\eq
\label{eq:b0sqbroken}
\expval{b_{0}^{2}} = \left(k^{2}-1 
+\frac{E}{K}\right)\frac{1}{\zeta'{}^{2}\aone^{2}}
\en
where $E(k)$ is the complete elliptic integral of the second kind.

In the unbroken phase, $\hat \phi=0$,  (\ref{eq:parametricequations})
implies
\eq
\label{eq:unbroken}
k^{2}=\frac12
\qquad
H= \frac{ 1}{8 \zeta'{}^{4} \aone^{4}}
\en
The solution is parametrized by $\aone$ in the range
\eq
\label{eq:kapparangeunbroken}
\expval{b_{0}^{2}} =\left(\frac{E}{K}-\frac12\right)\frac{ 
1}{\zeta'{}^{2}\aone^{2}} 
\ge \frac23
\en
In the broken phase,
\eqg
\label{eq:phisqbroken}
\hat\phi^{\dagger}\hat\phi  = 1 - \frac32 \expval{b_{0}^{2}}
=
1 - \frac32
\left(k^{2}-1 +\frac{E}{K}
\right)
\frac{1 }{\zeta'{}^{2}\aone^{2}}
\eng
Equation (\ref{eq:parametricequations}) gives $H$ and 
$\hat\phi^{\dagger}\hat\phi$ as functions of $k$ and $\aone$.
Equations (\ref{eq:phisqbroken}) and (\ref{eq:parametricequations}) 
combine to
give $\aone$ as a function of $k$.
\eqg
\label{eq:parambyk}
\zeta'{}^{2}\aone^{2} = \frac{4\lambda^{2}}{g ^{2} }(1-2k^{2}) 
+ \frac32
\left(k^{2}-1 +\frac{E}{K}
\right)
\qquad
0\le k^{2} \le \frac12
\eng
So $H$, $\hat\phi^{\dagger}\hat\phi$, and $\aone$ are functions of 
$k$.  
The solution in the broken phase is parametrized by $k$
in the range $0\le k^{2}\le 1/2$.

This is a classical solution of the equations of motion at leading
order in $\omegaz$.  The leading order solution should deform to an
exact classical solution order by order in $\omegaz^{-2}$.  This needs
to be verified.  And the semi-classical expansion around the classical
solution should be stable against small fluctuations.
This was shown for the Spin(4)-symmetric CGF in \cite{Friedan2022:Stability}.

\section{Energy-momentum tensor}

The CGF scalar and gauge energy-momentum tensors at leading order in 
$\omegaz$ are
\eqa
T^{\phi}{}^{\mu}_{\nu}
 &=
\frac{ \hbar  }{\az^{4}} \frac{1}{8\lambdaH^{2}}
\left[
\hat\phi^{\dagger}\hat\phi\, b_{0}^{2}
\left( 
3 e^{\mu}e_{\nu}-P^{\mu}_{\nu}
\right)
+
\left(\hat\phi^{\dagger}\hat\phi -  1 \right)^{2}
\left( 
e^{\mu}e_{\nu}-P^{\mu}_{\nu}
\right)
\right]
\\[2ex]
 T^{\gauge}{}^{\mu}_{\nu}  &=
\frac{\hbar }{\az^{4}}\frac1{2\gtwo ^{2}}
\left[\frac1{\aone^{2}}  \left(\frac{d b_{0}}{d\zeta}\right)^{2}
 +b_{0}^{4}\right] \left(3 u^{\mu}u_{\nu} +P^{\mu}_{\nu}
\right)
\ena
So the CGF energy-momentum tensor is that of a perfect fluid
\eqg
T^{\mu}_{\nu} =\rho u^{\mu}u_{\nu} + p P^{\mu}_{\nu}
\eng
Expressed in units of the microscopic energy density
\eq
\rhoz = \frac{m_{\Higgs}}{\az^{3}} = \frac{\hbar}{\az^{4}}
\en
the density $\rho$ is
\eqa
\frac{\rho }{\rhoz}
&= \frac3{2\gtwo ^{2}}
\left[\frac1{\aone^{2}} \left(\frac{d b_{0}}{d\zeta}\right)^{2}
 +b_{0}^{4}\right] 
+\frac{3}{8\lambdaH^{2}}\hat\phi^{\dagger}\hat\phi\, b_{0}^{2}
+\frac{1}{8\lambdaH^{2}} \left(\hat\phi^{\dagger}\hat\phi -  1 \right)^{2}
\\[1ex]
&= 
\frac{3}{\gtwo ^{2}}H
+\frac{1}{8\lambdaH^{2}} \left(\hat\phi^{\dagger}\hat\phi -  1 \right)^{2}
\ena
The pressure $p$ is
\eqa
\frac{p }{\rhoz}&= 
\frac1{2\gtwo ^{2}}
\left[\frac1{\aone^{2}} \left(\frac{d b_{0}}{d\zeta}\right)^{2}
 +b_{0}^{4}\right] 
-\frac{1}{8\lambdaH^{2}}\hat\phi^{\dagger}\hat\phi\, b_{0}^{2}
-\frac{1}{8\lambdaH^{2}} \left(\hat\phi^{\dagger}\hat\phi -  1 \right)^{2}
\\[1ex]
&= 
\frac{1}{\gtwo ^{2}}H
-\frac{1}{4\lambdaH^{2}}\hat\phi^{\dagger}\hat\phi\, b_{0}^{2}
-\frac{1}{8\lambdaH^{2}} \left(\hat\phi^{\dagger}\hat\phi -  1 \right)^{2}
\\[1ex]
&= 
\frac{1}{\gtwo ^{2}}H
-\frac{1}{4\lambdaH^{2}}\hat\phi^{\dagger}\hat\phi\,\expval{b_{0}^{2}}
-\frac{1}{8\lambdaH^{2}} \left(\hat\phi^{\dagger}\hat\phi -  1 \right)^{2}
\ena
where the rapidly oscillating term $b_{0}^{2}$  is replaced by its time 
average in the last step.

In the unbroken phase where $\phi =0$,
\eq
\frac{\rho }{\rhoz} = 
\frac3{\gtwo ^{2}}H
+\frac{1}{8\lambdaH^{2}}
\qquad
\frac{p}{\rhoz}=
\frac1{\gtwo ^{2}}H
-\frac{1}{8\lambdaH^{2}}
\en
with $H$ given by equation (\ref{eq:unbroken}).
The fluid is parametrized by $\aone$
in the range (\ref{eq:kapparangeunbroken}).

In the broken phase,
\eq
\frac{\rho }{\rhoz} = 
\frac3{\gtwo ^{2}}H
+\frac{9}{32\lambdaH^{2}}\expval{b_{0}^{2}}^{2}
\qquad
\frac{p }{\rhoz}=
\frac1{\gtwo ^{2}}H
- \frac{\hat\phi^{\dagger}\hat\phi}{4\lambda^{2}}\expval{b_{0}^{2}}
-\frac{9}{32\lambdaH^{2}}\expval{b_{0}^{2}}^{2}
\en
$H$ and $\hat\phi^{\dagger}\hat\phi$ are given by equation 
(\ref{eq:parametricequations})
and $\expval{b_{0}^{2}}$ 
is given by equation (\ref{eq:b0sqbroken}).
Equation (\ref{eq:parambyk})
parametrizes all three by $k$ 
in the range $0\le k^{2}\le 1/2$.

The equation of state 
is the same as calculated in the CGF cosmology \cite{Friedan2022:Atheory}.
The parametrization of $\rho$ and $p$ by $\aone$ in the unbroken 
phase and by $k$ in the broken phase
is the same as in the CGF cosmology.

\section{Irrotationality and the CGF rest frame}

The CGF is irrotational (at leading order in $\omegaz$)
because
$u_{\mu} = -\aone\partial_{\mu}T$ so
$\partial_{[\nu}u_{\rho]} 
= (u_{[\rho} \partial_{\nu]} \aone) /\aone$ so
\eq
u_{[\mu}\partial_{\nu} u_{\rho]} = 0
\en
Conversely, if $u^{\mu}(x)$ is an arbitrary irrotational four-velocity field then 
there exists a time coordinate $T(x)$ such that 
$P_{\mu}^{\alpha}\partial_{\alpha} T =0$,
i.e. $\partial_{\mu} T = -u_{\mu} (u^{\alpha}\partial_{\alpha} T)$.

Space-time is the union of the space-like hypersurfaces
parametrized by $T$.
\eq
M^{(3)}_{t} = \left\{ x \colon \; T(x) = t\right\}
\en
The hypersurfaces $M^{(3)}_{t}$ are orthogonal to $u_{\mu}$,
\eq
 v^{\mu} u_{\mu} = 0  \quad \Longleftrightarrow \quad
v^{\mu}\partial_{\mu}T =0
\en
The flow lines of the vector field $u^{\mu}\partial_{\mu}$
identify the hypersurfaces $M^{(3)}_{t}$ with each other.
So space-time is parametrized as $\Reals \times M^{(3)}_{0}$ with 
coordinates $x^{\mu} = (T, x^{i})$.
The four-velocity and metric are
\eqg
u^{\alpha} = (\aone^{-1},0,0,0)
\qquad
g_{00} = -\aone^{2}
\qquad
g_{0i} = 0
\\[1ex]
ds^{2}= \aM^{2} g_{\mu\nu}dx^{\mu}dx^{\nu} = \aM^{2}\aone(x)^{2}
\left[ -(dT)^{2} + g^{(3)}_{ij}(x) dx^{i}dx^{j}
\right]
\eng
This is the CGF rest frame.

\section{Adiabatic time evolution implies a continuity equation}

In the rest frame the leading order action of the CGF is
\eqg
\frac1\hbar S_{\CGF}
=
\int d \zeta\,\int d^{3}x  
\sqrt{\det g^{(3)}_{ij}}\;
\frac{3\omegaz^{3}}{g ^{2}}
\left[- \frac{1}{2} \left( \frac{d F}{d\zeta}\right)^{2}
+ A_{1} F^{2}
+ \frac{1}{2}F^{4}
+A_{0}
\right]    
\\[1ex]
A_{1} = \frac{g ^{2}\hat\phi^{\dagger}\hat\phi\aone^{2}}{8\lambda^{2}}
\qquad
A_{0}= \frac{g ^{2}(\hat\phi^{\dagger}\hat\phi-1)^{2}\aone^{4}}{24\lambda^{2}} 
\eng
Each volume element $d^{3}x$ of the CGF
is an independent anharmonic oscillator.
The oscillator coupling constants vary slowly in time compared to the 
period of oscillation.
In such an adiabatic time evolution,
the adiabatic invariant
$\oint p dq$ stays constant in time,
where $q$ is the oscillator degree of freedom,
$p$  its canonical conjugate,
and the integral is over one period of oscillation.
The conjugate variable to $q=F$ is
\eq
p =   d^{3}x \,\sqrt{\det g^{(3)}_{ij}}\;\hbar  
\frac{3\omegaz^{3}}{g ^{2}}
\frac{d F}{d\zeta}
\en
so the adiabatic invariant (measured in quanta) is
\eq
\frac1{2\pi\hbar}\oint p dq
=d^{3}x\,\sqrt{\det g^{(3)}_{ij}}\;N(x)
\en
\eqa
N(x) &=
\frac{3\omegaz^{3}}{2\pi g ^{2}}
\int_{0}^{2\pi} \left(\frac{d F}{d\zeta} \right)^{2} d\zeta
= \frac{\omegaz^{3}}{g ^{2}\zeta'{}^{2}}
\left[
1-k^{2}+(2k^{2}-1)\frac{E}{K}
\right]
\ena
Its constancy in time is the equation
\eq
0 = \frac{\partial}{\partial T}
\left(
 \sqrt{\det g^{(3)}_{ij}}\; N
 \right)
\en
Written covariantly this is the continuity equation
for a conserved number density $n(x)$.
\eq
n(x) = \frac{N(x)}{(\aM \aone)^{3}} 
\qquad
J^{\mu}(x) =  n(x)\, \aM^{3}\, u^{\mu}(x)\,\sqrt{-g}\,d^{4}x
\qquad
\partial_{\mu}J^{\mu}(x)  = 0
\en
$n(x)$ is the {density of quanta}.

\section{Summary of the parametrization}

Define functions of the elliptic parameter $k^{2}$,
\eqg
C_{1} =\frac12 { k^{2}(1-k^{2}) }
\qquad
C_{2} = 1-2k^{2}
\\[1ex]
C_{3} =k^{2}-1 +\frac{E}{K}
\qquad
C_{4} = \zeta' \left[1-k^{2}+(2k^{2}-1)\frac{E}{K}\right]
\qquad
\zeta' = \frac{ 2\pi}{4K}
\eng
The density, pressure, and number density are functions of two 
variables, $k$ and $\hat a=  {\zeta'}\aone$,
\eqa
\label{eq:parametrization}
\frac{\rho }{\rhoz}  &= 
\frac{1}{\hat a^{4}}
\frac{3}{\gtwo ^{2}}C_{1}
+\frac{1}{8\lambdaH^{2}} \left(\frac{4\lambda^{2}}{g ^{2}}\frac{C_{2}}{\hat a^{2}} -  1 \right)^{2}
\\[1ex]
\frac{p }{\rhoz}  &=
\frac{ 1}{\hat a^{4}}
\frac1{\gtwo ^{2}}
(C_{1}- C_{2}C_{3})
-\frac{1}{8\lambdaH^{2}} \left(\frac{4\lambda^{2}}{g ^{2}}\frac{C_{2}}{\hat a^{2}} -  1 \right)^{2}
\\[1ex]
\az^{3} n &=
\frac{ 1}{\hat a^{3}}
\frac{1}{g^{2}}
C_{4}
\ena
The fluid is parametrized by a one-dimensional subset of this 
two-parameter space,
\eq
\begin{array}{r@{\quad}c@{\qquad}l}
\text{unbroken phase} \colon
&
k^{2}=\frac12
&
0 < \hat a^{2} \le \frac 32 C_{3}
\\[2ex]
\text{broken phase} \colon
&
0\le k^{2} \le \frac12
&
\hat a^{2} =
\frac{4\lambda^{2}}{g ^{2} }C_{2}
+ \frac32 C_{3}
\end{array}
\en
The unbroken phase is parametrized by $\hat a(x)$,
the broken phase by $k(x)$.

The parametrization and therefore the equation of state
is the same as in the cosmological construction.

\section{Low density regime}
In the limit $k(x)\rightarrow 0$,
\eqa
K = \frac\pi2 \left(
1 + \frac{k^{2}}4 +\frac{9k^{4}}{64}  \right)+O(k^{6})
\qquad
E = \frac\pi2 \left(
1 - \frac{k^{2}}4  -\frac{3k^{4}}{64} \right)+O(k^{6}) 
\ena
from which, at leading order in $k^{2}$,
\eqg
\label{eq:smallk}
\zeta' =1
\qquad
\hat a^{2} = \aone^{2} = \frac{4\lambda^{2}}{g^{2}}
\qquad
ds^{2} = \aM^{2}\frac{4\lambda^{2}}{g^{2}}
\left[ -(dT)^{2} + g^{(3)}_{ij}(x) dx^{i}dx^{j}
\right]
\\[2ex]
\frac{\rho}{\rhoz} = \frac{3 g^{2} }{32\lambda^{4}}k^{2}
= 0.600 k^{2}
\qquad
\frac{p}{\rhoz} =
\frac{9}{256}
\frac{\gtwo ^{4}}{\lambda^{4}}
\left(
\frac1{g^{2}}
-\frac{1}{8\lambdaH^{2}}
\right)
k^{4}
\qquad
n=\frac{ 1 }{\az^{3}}
\frac{3}{16} 
\frac{g}{\lambda^{3}}
k^{2}
\eng
Small $k$ is the low density regime
${\rho}/{\rhoz} \ll 1$.
The low density equation of state is
\eqg
p= \frac{c_{a}}2  \frac{\rho^{2}}{\rhoz}
\qquad
c_{a}= 
\frac{\lambda^{2}\left(
8\lambdaH^{2}
-g^{2}
\right)
}{ g ^{2} }
=0.992
\eng
The density of quanta is related to the energy density by
\eqg
n = \frac{2\lambda}{g } \frac{\rho}{m_{\Higgs}} 
=\frac{\rho}{m_{W}}
\eng
so each quantum of the fluid has energy $m_{W}$.
A fluid mass $M$ consists of
$M/M_{W}$ quanta.

The average dark matter density  at the present time is
\eq
\expval{\rho_{\CDM}}_{\now}
= \Omega_{\CDM} \,\rho_{c}
= 2.3\ntimes 10^{-27} \kgunit/\munit^{3}
=  
4 \ntimes 10^{-56}\,
\rhoz
\en
where
$\rho_{c} = {3 H_{0}^{2}}/{8\pi G}$
is the critical density
and $\Omega_{\CDM} = 0.27$.  By (\ref{eq:smallk}),
\eq
\expval{k^{2}}_{\now} = 
\frac{32\lambda^{4}}{3 g ^{2} } \frac{\expval{\rho_{\CDM}}_{\now}}{\rhoz}
= 7 \ntimes 10^{-56}
\en
well within the low density regime.

The TOV stellar structure equations were solved in \cite{Friedan2022:DMStars}
using the equation of state given by (\ref{eq:parametrization}).
Figure~\ref{fig:MR} shows
the solutions as a spiral curve
in the radius-mass plane.
The curve is parametrized by the central density
$\rho_{\central}$ starting from $\rho_{\central}=0$, $M=0$.
A result from the 1960's 
says that the solutions are unstable against radial pulsations
except on the curve segment going from $M=0$ to
the maximum mass \cite[section 4.2.2 and references therein]{Thorne1967TheGR}.
The maximum mass is $M_{\max} = 9.15\ntimes 10^{-6}\Msun$
with central density $\rho_{\max} = 0.799 \rho_{0}$
corresponding to $k^{2}_{\max}=0.341$.
This is in the broken phase but not in the low density regime.
The TOV solutions can be stable only
in the range $0 < k^{2}_{\central} < k^{2}_{\max}$.
Presumably all of these solutions actually are stable.

Microlensing observations imply that,
if the halo dark matter is in the form
of such compact objects,
almost all the dark matter
is in  objects of mass less than $10^{-11}  \Msun$ \cite{Niikura:2017zjd}.
The star mass in the low density regime is $M=5 \ntimes 10^{-5} 
\Msun \,k^{2}_{\central}$
so $M=10^{-11}  \Msun$
corresponds to $k^{2}_{\central}= 2\ntimes 10^{-7} $.
So almost all of the CGF dark matter fluid should be in the low density regime.

\section{Possibilities of detection and verification?}

To find ways to detect the CGF,
it will be necessary to calculate
subleading effects ---
the interactions between the SM fields and the CGF
---
and to calculate
the mass spectrum of CGF stars in the halos and in the local region.
\begin{enumerate}
\item 
If the dark matter halo is composed of dark matter stars,
microlensing constrains their masses to be small.
The small mass CGF stars all have diameter
27\,cm independent of mass.
21\,cm radiation will scatter off such stars.
The halo might glow slightly in 21\,cm radiation
with a brightness depending on 
the interaction of the CGF with the electromagnetic field
and on the mass spectrum of CGF stars.

\item 
The CGF oscillation frequency in proper time is $m_{W}/\hbar$.
An electron-positron collider sitting near energy $m_{W}$ (or $2 m_{W}$)
might see  a resonance effect
when a CGF star passes through the interaction region,
depending on the strength of the interactions and the local 
mass spectrum of CGF stars.
\end{enumerate}
The CGF must satisfy the constraints of galaxy formation.
\begin{enumerate}
\item
The initial fluctuations of the CGF
have to be calculated,
along the lines of \cite{Friedan:2020poe,Friedan2022:Atheory,Friedan2022:Stability}.
The CGF fluctuations should satisfy 
the basic constraint
on the initial spectrum of dark matter fluctuations.
\item The time evolution of the initial CGF fluctuations
is to be calculated
in the classical CGF
dark matter universe.
Ordinary matter is a subleading correction.
Can subleading interactions radiate binding energy
during gravitational collapse?
Do the CGF fluctuations evolve in time
to form galactic halos?  of dark matter stars?  with what mass 
spectrum?  What is the local mass spectrum?
Might the irrotationality of CGF dark matter
explain the spherical shape of the dark matter halos?

\end{enumerate}
\vskip2ex
\phantomsection
\section*{Acknowledgments}
\addcontentsline{toc}{section}{\numberline{}Acknowledgments}
This work was supported by the Rutgers New High Energy Theory Center.
I thank C. Keeton for advice on microlensing and for pointing out \cite{Niikura:2017zjd}.

\vspace*{-1.5ex}

\bibliographystyle{ytphys}
\raggedright
\phantomsection
\addcontentsline{toc}{section}{\numberline{}References}
\bibliography{Literature}

\end{document}